\documentclass[aps,prl,superscriptaddress,reprint]{revtex4-1}

\usepackage[usenames, dvipsnames]{color}

\usepackage{graphicx}
\usepackage{xcolor}
\usepackage{rotating}
\usepackage{bm,amsmath,amssymb}
\usepackage[utf8]{inputenc}
\usepackage{footmisc}

\DeclareMathAlphabet{\mathpzc}{OT1}{pzc}{m}{it}

\usepackage{setspace}

\usepackage{slashed}
\usepackage{upgreek}

\newcommand{\nb}[1]{\color{blue}}

\usepackage[
	  pagebackref=false,
	  colorlinks=true,
      linkcolor=blue,
      urlcolor=blue,
      filecolor=black,
      citecolor=red,
      pdfstartview=FitV,
      pdftitle={},
        pdfauthor={},
        pdfsubject={},
        pdfkeywords={},
        pdfpagemode=None,
        bookmarksopen=true
      ]{hyperref}

\usepackage{graphicx}
\graphicspath{ {./images/} }

\usepackage{comment}

\usepackage[normalem]{ulem}
\usepackage{amsmath}
\usepackage{enumerate}
\usepackage{amsfonts}
\usepackage{epsfig}
\usepackage{mathbbol}

\def\Im{\mathop{\rm Im} }

\newcommand\half{{\ensuremath{\frac{1}{2}}}}
\newcommand\p{\ensuremath{\partial}}
\newcommand\pp{\ensuremath{\bm\nabla}}

\newcommand\vev[1]{{\ensuremath{\left\langle{#1}\right\rangle}}}

\newcommand{\be}{\begin{equation}}
\newcommand{\ee}{\end{equation}}
\newcommand{\bea}{\begin{eqnarray}}
\newcommand{\eea}{\end{eqnarray}}
\newcommand{\bega}{\begin{gather}}
\newcommand{\eega}{\end{gather}}

\newcommand{\bi}{\begin{itemize}}
\newcommand{\ei}{\end{itemize}}
\newcommand{\ben}{\begin{enumerate}}
\newcommand{\een}{\end{enumerate}}
\newcommand{\bca}{\begin{cases}}
\newcommand{\eca}{\end{cases}}
\newcommand{\bln}{\begin{align}}
\newcommand{\eln}{\end{align}}
\newcommand{\bst}{\begin{split}}
\newcommand{\est}{\end{split}}
\def\ie{\begin{equation}\begin{aligned}}
\def\fe{\end{aligned}\end{equation}}
\newcommand{\bma}{\le(\begin{matrix}}
\newcommand{\ema}{\end{matrix}\ri)}
\newcommand{\bwt}{\begin{widetext}}
\newcommand{\ewt}{\end{widetext}}

\newcommand\lam{\lambda}
\newcommand\Lam{\Lambda}

\newcommand\Th{{\Theta}}
\def\th{{\theta}}

\newcommand\ov{\over}
\newcommand\ha{{\half}}

\def\le{\left}
\def\ri{\right}

\newcommand\sA{{\ensuremath{{\mathcal A}}}}

\DeclareMathAlphabet{\pazocal}{OMS}{zplm}{m}{n}

\newcommand{\eps}{{\epsilon}}

\begin{document}

\title{ Chimeric states of matter: Meissner effect without superconductivity}

\author{ Michael J. Landry}
\email{mjlandry@mit.edu}
\affiliation{Quantum Measurement Group, MIT, Cambridge, MA 02139, USA}
\affiliation{Department of Nuclear Science and Engineering, MIT, Cambridge, MA 02139, USA}
\affiliation{Department of Physics, MIT, Cambridge, MA 02139, USA}

\author{Mingda Li}
\affiliation{Quantum Measurement Group, MIT, Cambridge, MA 02139, USA}
\affiliation{Department of Nuclear Science and Engineering, MIT, Cambridge, MA 02139, USA}

\begin{abstract}

Symmetry is central to how we classify phases of matter: solids break spatial translations, superfluids break particle-number conservation, and superconductors ``break'' gauge symmetry. Mixed anomalies involving higher-form symmetries, however, present a generalization of spontaneous symmetry breaking that admits a wider and more versatile set of possibilities. We introduce chimeric states of matter, in which aspects of broken and unbroken phases coexist. We find that the Meissner effect---usually regarded as the defining hallmark of superconductivity---can occur in media that are resistive or even insulating when probed by electric fields. We demonstrate this by constructing an effective field theory of “symmetry chimerization” and propose that Josephson junction networks could provide a laboratory realization. These results broaden the landscape of possible phases of matter, showing that physical media can mix features of symmetry-restored and symmetry-broken states in a single substrate.

\end{abstract}

\maketitle

Symmetry guides our understanding of phases of matter: crystals break translational symmetry, superfluids break particle-number conservation, and superconductors are often described as ``breaking'' gauge symmetry through the Anderson-Higgs mechanism. In each case, the broken symmetry gives rise to Goldstone modes---low-energy gapless excitations that reflect the continuous symmetry that the ground state no longer respects. Phonons in crystals and phase fluctuations in superfluids are familiar examples, while in superconductors, the would-be Goldstone is absorbed by the gauge field, giving rise to a massive photon. Fluids, by contrast, preserve all internal symmetries other than boosts, and thus host no independent Goldstone excitations. This Landau picture and its effective field theory descendants remain a dominant organizing language for many-body physics \cite{Son:2002zn,Baggioli:2021ntj,Landry:2020ire,Nicolis:2011cs,Dubovsky:2011sj,Weinberg:1996kr,Nicolis:2015sra,Landry:2019iel}.

\begin{figure}[!htbp]
  \centering
  \includegraphics[width=0.45\textwidth]{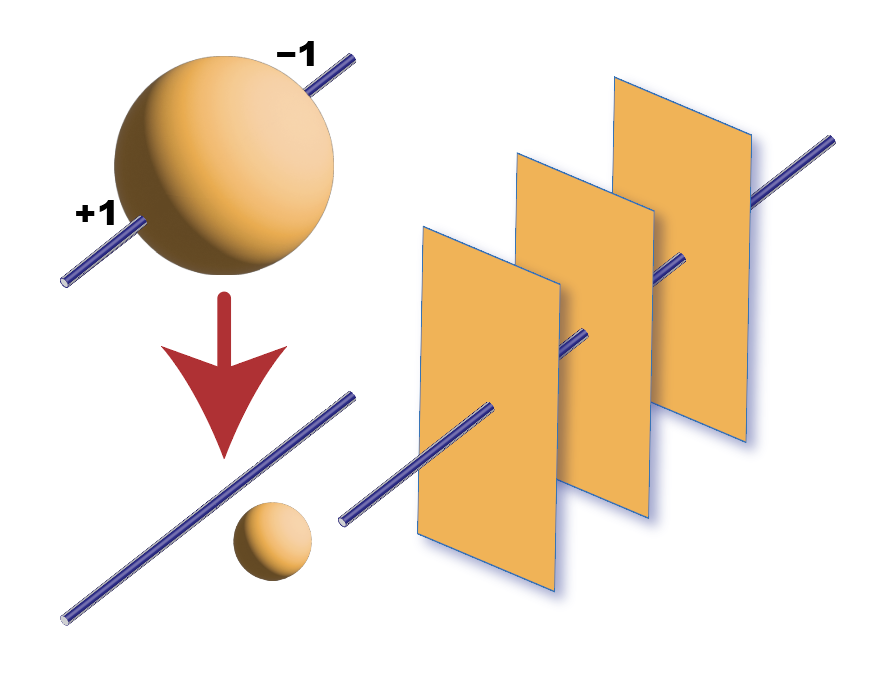}
  \caption{Infinitely extending lines intersect winding planes. Left: Spherical winding plane. Sum of {\it oriented} intersections is zero ($+1$ into, $-1$ out of surface). Spherical winding planes can contract to zero without changing the intersection number. Right: Stack of infinitely extending winding planes. Total oriented intersection is non-zero but cannot change unless planes are punctured or torn. Conclusion: intersection number is topologically protected and hence conserved. } \label{fig:1}
\end{figure}

In recent years, that taxonomy has expanded dramatically. Topological phases~\cite{Witten:2015aoa,Hasan:2010xy,Qi:2010qag,Fu:2006djh}, higher-form symmetries~\cite{Gaiotto:2014kfa,Glorioso:2018kcp,Armas:2018zbe,Hofman:2018lfz,Grozdanov:2016tdf,Son:2009tf,Landry:2021kko}, non-invertible symmetries~\cite{Das:2022fho,Choi:2022jqy,Cordova:2022ieu,Koide:2021zxj,Choi:2022zal}, and anomalies~\cite{Landry:2022nog,Glorioso:2017lcn,Delacretaz:2019brr} have revealed new organizing principles beyond conventional symmetry breaking. Yet even within these generalized frameworks, one  implicit assumption remains: symmetries are either fully preserved or fully broken. 

In this Letter,  we show that this dichotomy is incomplete: phases can interpolate, combining features of both. We refer to these as chimeric states of matter, which emerge naturally when the concept of spontaneous symmetry breaking (SSB) is generalized within the framework of higher-form symmetry.

The spontaneous breaking of a continuous Abelian symmetry results in a shift-symmetric Goldstone boson $\phi$. Surfaces of constant $\phi$, known as winding planes, are conserved: the oriented intersection number between any such plane and an infinitely extended line remains constant in time (see Fig.~\ref{fig:1}). By Noether’s theorem, this conservation must originate from an underlying symmetry. Because the conserved quantities are extended rather than point-like, the concept of symmetry must be generalized; the conservation of $p$-dimensional objects arises from a so-called ``$p$-form symmetry.'' In this case, the symmetry associated with conserved winding planes is a 2-form symmetry. When the broken symmetry is gauged, these winding planes are no longer gauge-invariant observables, and the 2-form symmetry is destroyed. In this sense, a {\it mixed anomaly} emerges between the broken symmetry and the 2-form symmetry.  

\begin{figure}[htbp]
  \centering
  \includegraphics[width=0.45\textwidth]{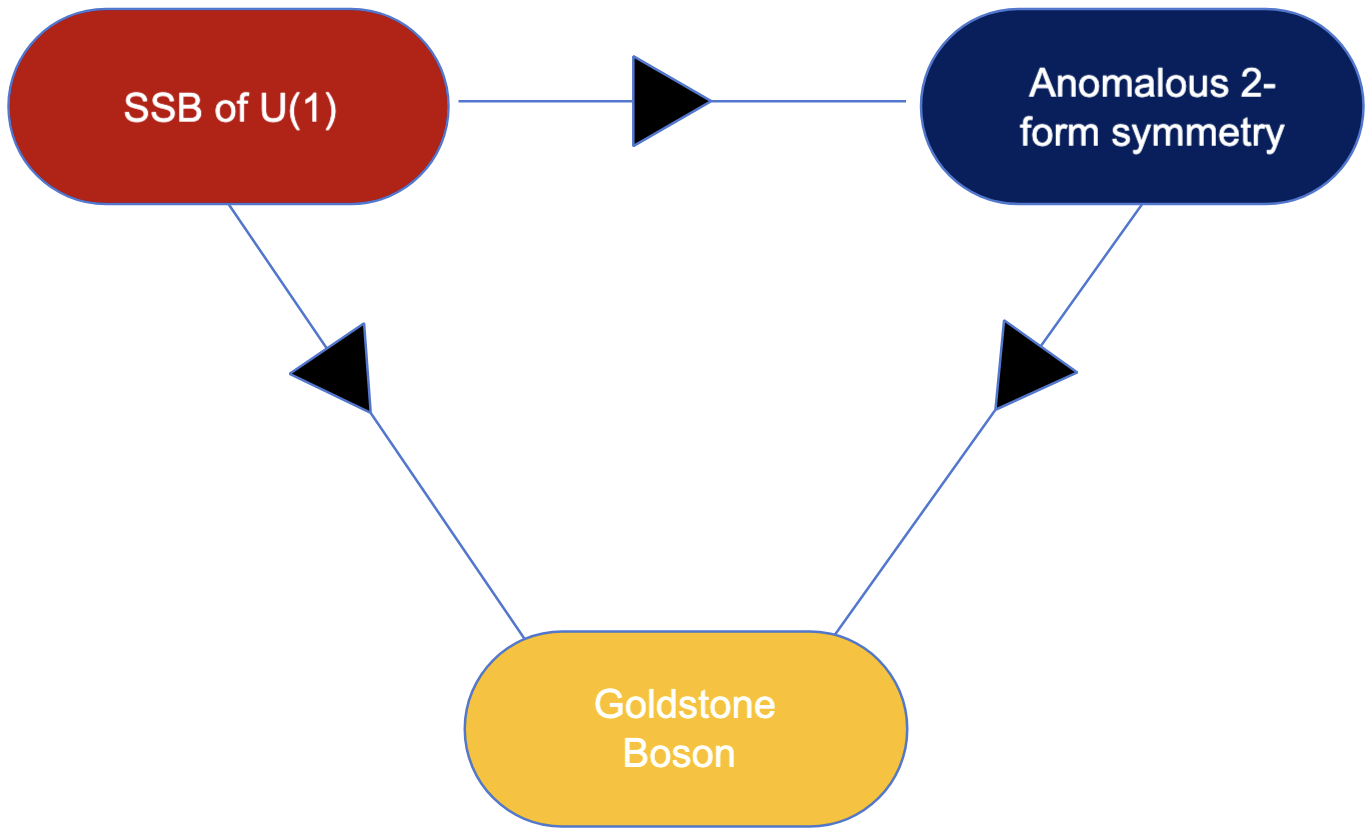}
  \caption{ The anomalous 2-form symmetry picture generalizes the notion of SSB of the $U(1)$ symmetry and yields Goldstone bosons. SSB implies mixed anomaly, but not necessarily vice versa.}\label{fig:2}
\end{figure}

Every SSB of a continuous Abelian symmetry implies the existence of such a mixed anomaly, though the reverse is not necessarily true (see Fig.~\ref{fig:2}). Nevertheless, both SSB and mixed anomalies with 2-form symmetries are sufficient to guarantee the existence of a Goldstone mode. This mixed-anomaly viewpoint generalizes the usual Goldstone logic \cite{pich2020goldstone,naegels2021goldstone} and yields new phenomenology. Recasting superconductivity in these terms reveals ``partial symmetry realizations,'' states that retain part of the anomaly structure while violating other aspects. The result is a family of what we call chimeric responses: systems that can, for example, exhibit a full Meissner effect yet still have non-zero d.c. resistivity, forming {\it chimeric conductors}. Similarly, for global $U(1)$ symmetry, we identify a {\it chimeric superfluid}, a phase with vanishing superfluid order parameter but a topological remnant of phase stiffness.

To connect with experiment, we present a simple Josephson-junction-network lattice model that reproduces the electromagnetic responses of the effective theory and offers a route to realization. Our results show that symmetry breaking need not be all-or-nothing, revealing a richer landscape of phases that merge traits once thought mutually exclusive.

{\it Mixed anomalies:} In the presence of SSB for a global $U(1)$, there are two conserved currents of interest. The broken $U(1)$ symmetry has a conserved Noether current $J^\mu=(\rho,\bm J)^\mu$, which describes particle-number conservation in superfluids. This current exists in both broken and unbroken phases, and satisfies the conservation equation
\be
     \p_\mu J^\mu = 0 ,\quad \iff \quad \p_t \rho + \pp\cdot\bm J = 0. 
\ee
When this symmetry is broken, there exists Goldstone $\phi$. 
Using this Goldstone and 4D Levi-Civita tensor $\eps^{\mu\nu\lam\rho}$, we can construct the 3-form current 
$K^{\lam\mu\nu} = \eps^{\lam\mu\nu\rho}\p_\rho \phi$,
which is conserved in the absence of vortical defects:
\be
     \p_\lam K^{\lam\mu\nu} = 0. 
\ee
The conservation of this 3-form current signals an emergent 2-form symmetry. 

Crucially, this 2-form symmetry is not independent of particle-number conservation. If we couple the system to a background gauge field $A_\mu$, the conservation law is destroyed, indicating a {\it mixed anomaly}:
\be\label{super}
     K^{\lam\mu\nu} = \eps^{\lam\mu\nu\rho}(A_\rho+\p_\rho \phi), \quad \p_\lam K^{\lam\mu\nu} = \tilde F^{\mu\nu},
\ee
with $\tilde F^{\mu\nu}=\frac{1}{2}\,\epsilon^{\mu\nu\rho\sigma} F_{\rho\sigma}$ the dual electromagnetic field strength.

Recent results show that the existence of such a mixed anomaly is itself sufficient to guarantee the presence of propagating gapless modes, generalizing Goldstone’s theorem~\cite{Delacretaz:2019brr}. In light of this, the anomaly-based picture naturally generalizes the symmetry-breaking picture. 

Suppose we violate the anomaly structure by introducing sources $\Gamma^{\mu\nu}$:
\be
     \p_\lam K^{\lam\mu\nu} = \tilde F^{\mu\nu} + \Gamma^{\mu\nu}.
\ee
Then the system moves toward the unbroken phase. 
With $\Gamma^{0i}$ the vortex density and $\Gamma^{ij}$ the vortex current, we obtain a familiar result, that vortex proliferation destroys superconducting phase. 

\begin{table*}[htbp]
\begin{center}
\caption{Classification of states according to chimeric anomalies for global and local $U(1)$. }\label{T}
\begin{tabular}{|c || c | c ||} 
 \hline
  & Global $U(1)$ ($F_{\mu\nu}=0$) & Local $U(1)$ (dynamical $F_{\mu\nu}$)  \\ [0.5ex] 
 \hline\hline
 $\Gamma^{0i}=0$, $\Gamma^{ij} = 0$ & Superfluid & Superconductor  \\ 
 \hline
 $\Gamma^{0i}\neq 0$, $\Gamma^{ij} \neq  0$ & Unbroken phase (e.g. fluid) & Ordinary conductor/insulator  \\
 \hline
 $\Gamma^{0i} \neq 0$, $\Gamma^{ij} = 0$ & Superfluid vortex glass & Superconducting vortex glass \\
 \hline
 $\Gamma^{0i} = 0$, $\Gamma^{ij} \neq  0$ & Chimeric superfluid & Chimeric conductor/insulator  \\
 \hline
\end{tabular}
\end{center}
\end{table*}

{\it Symmetry chimerization:} Consider superconductors, where the relevant $U(1)$ symmetry is gauged. While superfluids and superconductors have very different microscopic origins, their long-distance physics admit similar mathematical descriptions; see Table~\ref{T}. Now the Goldstone boson is absorbed by the dynamical gauge field $A_\mu$, and the anomaly relation above captures the superconducting phase, while its violation corresponds to the ordinary conducting phase. Furthermore, the conserved current $J^\mu=(\rho,\bm J)^\mu$ describes the electromagnetic charge and current densities.

The anomaly need not be all-or-nothing. In the absence of boost symmetry, some components of $K^{\lam\mu\nu}$ may obey the anomaly relation, while others do not. This possibility leads to new phases that lie between the familiar categories.

When boost symmetry is broken, several scenarios emerge. The anomaly for superconductors is exact $(\Gamma^{\mu\nu} = 0)$, typically indicating Higgs phase. However, the anomaly picture is more general, allowing the possibility of ``Higgsless superconductors''~\cite{Diamantini_2015}. (Note: type-II superconductors admit only {\it gapped} vortices, so their homogeneous equilibrium state is vortex-free.) 
In ordinary conductors and insulators, all components of the anomaly are violated, meaning that vortices can appear freely ($\Gamma^{\mu\nu}$ is fully non-zero). 
Vortex glasses exhibit pinned vortex density $(\Gamma^{0i} \neq 0,\Gamma^{ij}=0)$. These states share certain aspects of both superconducting and normal phases~\cite{FisherDanielS.1991Tfqd,nattermann2000vortex}. 
We define chimeric conductors and insulators to have vanishing vortex density but non-zero vortex current. In this case, 
\be
     \p_\mu K^{\mu ij} = \tilde F^{ij} + \Gamma^{ij} ,\quad \p_\mu K^{\mu 0i} = \tilde F^{0i}. 
\ee
These are the main focus of this work. Remarkably, they can exhibit the Meissner effect---complete expulsion of magnetic fields---while still showing non-zero d.c. resistivity. This anomaly-based perspective softens the sharpness of phase distinctions: the degree of anomaly violation can be tuned continuously through $\Gamma^{\mu\nu}$.

Even richer possibilities arise if rotational symmetry is broken. In principle, a material could behave as a superconductor for fields aligned along one axis, as an insulator along another, and as a chimeric conductor along a third. Exploring such anisotropic chimeric states remains an exciting direction for future work.

{\it Chimeric conductors and insulators:} 
The field contents are the electromagnetic gauge field $A_\mu$ and a 2-form field $b_{\mu\nu}=b_{\nu\mu}$ (see End Matter). It is convenient to define
     $b_i = b_{0i},$ and $ c_i = \ha \eps_{ijk} b_{jk} $. 
The identically-conserved electromagnetic current can be expressed in terms of this 2-form field by
\be
     J^\mu = \ha \eps^{\mu\nu\lam\rho} \p_\nu b_{\lam\rho},\quad \p_\mu J^\mu \equiv 0. 
\ee
We may interpret $J^0 = \pp\cdot \bm c$ and $\bm J=\pp\times\bm b - \p_t\bm c$ as the {\it free} charge and current densities. Letting $\bm D = \eps \bm E$ and $\bm H= \bm B/\mu$, Maxwell's equations always hold,
\be\label{Maxwell:1}
    \bm \pp\times \bm H - \p_t \bm D = \bm J,\quad \pp\cdot\bm D = J^0. 
\ee
Depending on the particular phase of matter, additional equations describing the motion of charges are needed, which are furnished by an effective action (see End Matter). If $K^{\lam\mu\nu}$ is conserved (up to anomaly), the 2-form field $b_{\mu\nu}$ must appear in the effective action only in the package $J^\mu$. If, however, the $\mu=A,\nu=B$ component $\Gamma^{AB}$ is non-zero, then $b_{AB}$ may appear in the action outside of this package. For superconductors, neither $\bm b$ nor $\bm c$ may appear outside $J^\mu$, while for ordinary conductors and insulators both may. In vortex glasses, $\bm b$ but not $\bm c$ may appear outside $J^\mu$. 

Chimeric conductors and insulators are defined by the absence of vortex density but the presence of vortex currents, so $\bm c$ but not $\bm b$ may appear outside the package $J^\mu$. It is convenient to decompose the current into two components: $\bm J = \bm J_{\rm cond} + \bm J_{\rm Meis}$ with
     $\bm J_{\rm cond} = - \p_t\bm c,$ and $ \bm J_{\rm Meis} = \pp\times\bm b.$ 
The equations of motion yield Ohm's law and the {\it second} London equation with penetration depth~$\lam$,
\be
     \bm J_{\rm cond} = \sigma \bm E,
\quad
     \bm B = - \mu \lam^2 \pp\times \bm J.
\ee
Consequently, $\pp^2 \bm B = \lam^{-2} \bm B $.  Thus, chimeric conductors combine two hallmarks usually thought to be mutually exclusive: they conduct electricity dissipatively, while exhibiting full Meissner effect.

To describe chimeric insulators, send $\sigma\to 0$. 
The equations of motion are then Maxwell's equations supplemented by 
     $\bm B = -\mu\lam^2 \pp\times \bm J$ and $\pp\cdot\bm J=0$.  
The result is a state with zero electrical conductivity but full Meissner effect. 

{\it Lattice model:} To demonstrate the microscopic feasibility of chimeric states, consider a lattice model based on Josephson-coupled superconducting grains embedded in an insulating host. Each lattice site $i$ is characterized by: (1) a superconducting phase operator $\theta_i$ and (2) a Cooper-pair number operator $n_i$, with canonical commutator 
$[\theta_i,n_j]=i\hbar \delta_{ij}$. To couple to electromagnetism, we introduce the scalar potential $A_0$ and the vector potential $\bm A$. On the lattice we define link integrals
\be
     \sA_{ij} = {2e\ov\hbar} \int_{\bm x_i}^{\bm x_j}d\bm x\cdot\bm A.
\ee
Finally, we introduce an auxiliary link field $\xi_{ij}=-\xi_{ji}$, with conjugate momentum $\pi_{ij}$ satisfying 
     $[\xi_{ij},\pi_{kl}] = i\hbar \delta_{ik}\delta_{jl}.$ 
This link variable is crucial for symmetry chimerization.

With these definitions, the Hamiltonian is
\be\begin{split}
     H = &\sum_i {Q_i^2\ov 2C} - E_J \sum_{\vev{ij}} \cos\left(\theta_i-\theta_j -\sA_{ij} -\xi_{ij}\right) \\ &
     \sum_{\vev{ij}} {\pi^2_{ij}\ov 2M} - U \sum_p \cos\Phi_p - \sum_i Q_i  A_0(\bm x_i) \\ &+ H_{\rm Maxwell},
\end{split}\ee
where $\vev{ij}$ indicate nearest neighbor pairs. 
The first term is the charging energy with capacitance $C$ and charge $Q_i = 2en_i$. The second is the Josephson coupling written in the usual gauge-invariant form, but modified by $\xi_{ij}$. The third and fourth govern the dynamics of $\xi_{ij}$, where $\Phi_p$ is the oriented sum of $\xi_{ij}$ around plaquette $p$. The fifth term is the usual coupling of charge to the scalar potential. $H_{\rm Maxwell}$ is the standard lattice Hamiltonian for the free electromagnetic field. The Hamiltonian is invariant under gauge transformations
\bega
     \th_i\to \th_i + \chi_i(t),\quad \sA_{ij} \to \sA_{ij} + \chi_i(t)-\chi_j(t),\\ A_0(\bm x_i) \to A_0(\bm x_i) - \dot \chi_i(t) , 
\end{gather}
and, for time-independent $\kappa_i$, under shift symmetry
\be
     \th_i \to \th_i + \kappa_i ,\quad \xi_{ij} \to \xi_{ij} + \kappa_i - \kappa_j.
\ee

\begin{figure}[htbp]
  \centering
\includegraphics[width=0.4\textwidth]{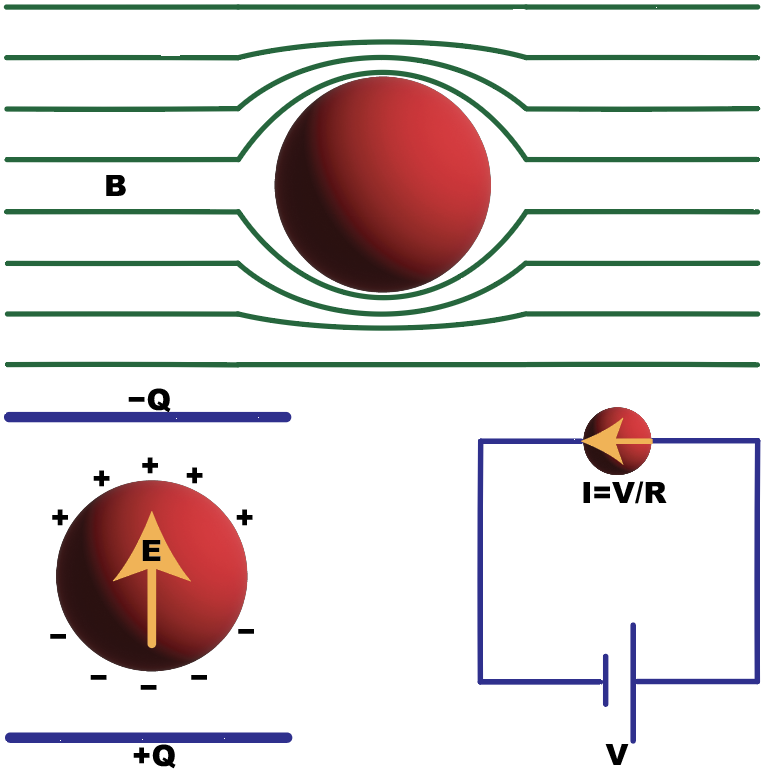}
  \caption{Top: chimeric conductors and insulators fully screen magnetic fields according to the Meissner effect. Bottom left: chimeric insulator only partially screens electric fields. Bottom right: chimeric conductor admits dissipative d.c. current.}\label{fig:3}
\end{figure}

The auxiliary field $\xi_{ij}$ controls whether the anomaly is chimerized or unbroken. At small $U$, $\xi_{ij}$  fluctuates freely and sources all components of the 3-form current. Such vortex proliferation leads to ordinary insulation. At large $U$, the curl of $\xi_{ij}$ is gapped (confining phase~\cite{Banks:1977cc}), so $\xi_{ij}$ becomes effectively curl-free in the infrared, $\xi_{ij} \sim \psi(\bm x_j)-\psi(\bm x_i)$. In this regime, only spatial components of the 3-form current are sourced, realizing the chimeric anomaly. 

Replacing the insulating host by a conductor in principle allows the same construction to produce a chimeric conductor rather than an insulator.

In the curl-free sector (large $U$), we write $\xi_{ij}\to {\hbar\ov 2e} \pp\psi$ in the continuum limit. Expanding the Josephson term to quadratic order, absorbing $\theta$ into $\psi$ in the gradient term, integrating out $\theta$, and performing a Legendre transform, the leading-order effective action becomes
\be\begin{split}\label{eq:chimins2}
     S_{\rm lat} = \int d^4 x \bigg( \ha {\eps} \bm E^2 - {1\ov 2\mu}\bm B^2 - {1\ov 2 \mu \lam^2} (\bm A -\pp \psi)^2\bigg),
\end{split}\ee
with magnetic penetration depth  
     $\lam = \sqrt{\hbar \ell / 4 e^2 \mu E_J}.$ 
While the above action possesses no 2-form fields, it turns out to be the mathematical dual of the 2-form description for chimeric insulators (see End Matter). Moreover, identifying the electrical current density with the variation of the action with respect to $\bm A$, the equations of motion are Maxwell's equations supplemented by  
\be
     \bm B = -\mu \lam^2 \pp\times\bm J,\quad \pp\cdot\bm J=0.
\ee
This result confirms that this is a lattice model for a chimeric insulator with penetration depth $\lam$.

{\it Chimerization of global $U(1)$:} 
Gauge symmetries, strictly speaking, can never be spontaneously broken: the Higgs phase is not characterized by a local order parameter but by the presence of a massive gauge boson. It is therefore useful to repeat the construction for a global $U(1)$ symmetry, where spontaneous symmetry breaking has an unambiguous meaning. We may ask: does a chimerized global symmetry corresponds to broken or unbroken $U(1)$? 

Setting $A_\mu=0$ in in the full equations for chimeric conductors (see End Matter) yields the equations of motion for chimerized global $U(1)$:
\bega
      \p_t J^0 = {\sigma\ov \chi}\pp^2 J^0    ,\quad \pp\times\bm J = 0,
\end{gather}
which yields no propagating mode. Hence Goldstone's theorem fails, so the symmetry is unbroken. A direct computation of the order parameter $\vev{e^{i\th}}$ confirms this conclusion (see End Matter). 

Nevertheless, this phase shares substantial qualitative similarities with a superfluid. The condition $\pp\times\bm J = 0$ admits curl-free stationary currents---for instance, a uniform current 
$\bm J = j_0\bm e_z$---even in equilibrium. Thus, the chimeric superfluid supports persistent, dissipationless  charge transport: charge can move coherently without a phase-coherent condensate. Table~\ref{T} compares states of matter with global and gauged $U(1)$. 

{\it Conclusion:} 
We have shown that the usual dichotomy between broken and unbroken symmetry can be relaxed: mixed anomaly structures permit phases that exhibit features of both. Focusing on chimerization of the electromagnetic $U(1)$ symmetry, we constructed coarse-grained effective theories and a microscopic lattice model that produce two new families of phases---chimeric conductors and chimeric insulators---which obey Ohm’s law or remain insulating under applied electric fields, yet expel magnetic flux according to the second London equation. The Meissner effect, therefore, need not be exclusive to superconductors. Beyond its conceptual implications, this expanded realization of flux expulsion hints at potential applications in contactless transport, magnetic shielding, and levitation technologies akin to maglev systems, where dissipationless magnetic responses could be achieved without full superconductivity.

{\it Acknowledgments:} MJL and ML acknowledge support from the National Science Foundation (NSF) Convergence Accelerator Award No.~2345084, and U.S. Department of Energy (DOE), Basic Energy Sciences (BES), Award No.~DE-SC0020148. 

\bibliography{references}

\section{End Matter}

{\it Mixed anomalies:} Suppose the $U(1)$ is global, with $A_\mu$ a background probe. Introduce external 3-form source $h_{\lam\mu\nu}$ for $K^{\lam\mu\nu}$. After integrating out the microscopic matter fields, the system is described by a generating functional 
$W[A_\mu,h_{\lam\mu\nu}],$ 
from which correlation functions follow by functional differentiation. In particular, 
     \be \hat J^\mu(x) = {\delta W\ov \delta A_\mu(x)},\quad \hat K^{\lam\mu\nu}(x) = {1\ov 3!} {\delta W\ov \delta h_{\lam\mu\nu}(x)},\ee  
define the on-shell electromagnetic current and the 3-form current, respectively.

If both symmetries are exact, conservation of these currents corresponds to gauge invariances of the generating functional.  
The presence of a mixed anomaly modifies this invariance in a controlled way:
\be\begin{split}
     W\left[A_\mu + \p_\mu \Lam,h_{\lam\mu\nu} + \p_{[\lam} \kappa_{\mu\nu]}\right] = W[A_\mu,h_{\lam\mu\nu}] \\
     + \ha \int d^4 x \, \eps^{\mu\nu\lam\rho} \p_\mu A_\nu \kappa_{\lam\rho} .
\end{split}\ee

To reproduce this structure in an EFT, promote the gauge variation of $h_{\lam\mu\nu}$ to a dynamical field $b_{\mu\nu}$. This antisymmetric tensor defines a conserved current, 
$J^\mu = \ha \eps^{\mu\nu\lam\rho} \p_\nu b_{\lam\rho}$. If the action depends on $b_{\mu\nu}$ only through the combination
$H_{\lam\mu\nu} = h_{\lam\mu\nu} + \p_{[\lam} b_{\mu\nu]}$, 
then gauge invariance of the 2-form symmetry is automatic, and $\hat K^{\lam\mu\nu}$ is conserved. This is known as the {\it Stückelberg trick}. If the above gauge symmetry fails to be satisfied, then $\hat K^{\lam\mu\nu}$ is not conserved. 

The minimal effective action realizing the anomaly is
$S = S_0[F_{\mu\nu},H_{\lam\mu\nu}] + \int d^4 x \, J^\mu A_\mu,$ 
where $F=dA$. The first term, $S_0$, is gauge-invariant under  
\be\label{2gauge}
     h_{\lam\mu\nu} = h_{\lam\mu\nu}+ \p_{[\lam} \kappa_{\mu\nu]},\quad b_{\mu\nu}\to b_{\mu\nu} - \kappa_{\mu\nu},
\ee
while the $J^\mu A_\mu$ term explicitly breaks invariance in precisely the way required to reproduce the mixed anomaly. Promoting $A_\mu$ to a dynamical field yields the superconducting phase. 

Working to quadratic order in fields, and turning off sources, $h=0$, the most general action is 
     \be S_{\rm sc} = \int d^4 x \left( \ha \eps \bm E^2 - {1\ov 2\mu}\bm B^2 + {1\ov 2\chi} (J^0)^2 
     - \ha \mu\lam^2 \bm J^2  + J^\mu A_\mu \right),\ee 
with $\eps$ is the electric permittivity, $\mu$ the magnetic permeability, $\lam$ the magnetic penetration depth, and $\chi$ the susceptibility. Notice that with $h=0$, $J^\mu  = {1\ov 3!} \eps^{\mu\nu\lam\rho}H_{\nu\lam\rho}$. The equations of motion for $A_\mu$ reproduce Maxwell's equations in matter and the 2-form dynamics yield 
      \be\bm E = \mu\lam^2 \p_t \bm J ,\quad \bm B = -\mu\lam^2 \pp\times \bm J.\ee

{\it Schwinger-Keldysh field theory:} Oftentimes, dissipation is crucial. Then it is necessary to apply the Schwinger-Keldysh (SK) approach to effective field theory. We review the basics here. 

For every field $A_\mu,b_{\mu\nu}$, there exist partner fields denoted by subscript $a$, which serve as generalized Lagrange multipliers that also encode information about statistical fluctuations: $A_{a\mu},b_{a\mu\nu}$. We will neglect temperature fluctuations and turn off background fields $h,h_a$. Our action $I_{\rm EFT}[A_\mu, b_{\mu\nu},A_{a\mu}, b_{a\mu\nu}]$ (we use $I$ as opposed to $S$ to indicate SK) is then constructed as the most general action consistent with symmetries and subject to the following conditions: (1) Impose three constraints derived from unitarity,
     \bega
          I^*_{\rm EFT}[A,A_a,b,b_a] = - I_{\rm EFT}[A,-A_a,b,-b_a],\\
          \Im \, I_{\rm EFT} \geq 0,\quad 
          I_{\rm EFT}[A,b,A_a=b_a=0] = 0. 
     \end{gather}
(2) Require invariance under the $\mathbb Z_2$ dynamical KMS transformations, $I_{\rm EFT}[A,A_a,b,b_a] = I_{\rm EFT}[\tilde A,\tilde A_a,\tilde b,\tilde b_a]$, where for some time-reversing symmetry $\Th$: 
\bega
\tilde A_\mu = \Th A_\mu,\quad \tilde A_{a\mu} = \Th A_{a\mu} + i\beta_0 \Th \p_t A_\mu,\\
\tilde b_{\mu\nu} = \Th b_{\mu\nu},\quad \tilde b_{a\mu\nu} = \Th b_{a\mu\nu}  + i\beta_0 \Th \p_t b_{\mu\nu}.
\end{gather}
     At quadratic order, it ensures the fluctuation-dissipation theorem is satisfied. 
     We assume all systems are parity and charge symmetric.

{\it Non-Stückelberg trick:} 
To formulate an action on the SK contour that enjoys a mixed anomaly, we have
\be\label{eq:I}
     I = I_0[A_\mu,H_{\lam\mu\nu},A_{a\mu},H_{a\lam\mu\nu}] + \int d^4 x \left(J^\mu A_{a\mu} + J_a^\mu A_\mu\right),
\ee
where $J_a^\mu$ is the $a$-type partner of $J^\mu$. The term $I_0$ is invariant under the two gauge transformations: $h_{\lam\mu\nu} \to h_{\lam\mu\nu}+ \p_{[\lam} \kappa_{\mu\nu]},$ $ b_{\mu\nu}\to b_{\mu\nu} - \kappa_{\mu\nu}$, and its corresponding $a$-type version. 
If we wish to construct a theory with full mixed anomaly (i.e. a superconductor) then our action must respect the above gauge symmetry. In this way~\eqref{eq:I} characterizes the general form of the action for a superconductor. If, however, we wish to violate the anomaly, say in the $A,B$ component, that is  
     $\Gamma^{\mu\nu} = \Gamma^{AB}\delta_A^\mu \delta_B^\nu $, 
then we must violate the gauge symmetry. In particular, we must allow $b_{AB},b_{aAB}$ to appear in the action outside of the gauge-invariant packages $H_{\lam\mu\nu},H_{a\lam\mu\nu}$. We call $b_{AB},b_{aAB}$ {\it non-Stückelberg fields}. 

The action for ordinary conductors must fully violate the mixed anomaly ($\Gamma^{\mu\nu}\neq 0$), so $\bm b,\bm b_a$ and $\bm c,\bm c_a$ are all non-Stückelberg fields. It turns out, however, that the correct prescription is to impose certain time-independent gauge symmetries~\cite{Landry:2019iel,Landry:2021kko,Landry:2022nog}, 
$\bm c \to \bm c+ \pp\times \bm v(\bm x)$
for time-independent vector $\bm v(\bm x)$. 

Decompose the action into (anomalous) Stückelberg and non-Stückelberg sectors~\cite{landry2023activeactionseffectivefield}: $I_{\rm oc} = I^{\text{(Stück)}} + I_{\rm oc}^{\rm (non)}.$ The (anomalous) Stückelberg terms will be the same for all phases, and hence take the form of the superconductor action on the SK contour. 
The non-Stückelberg terms encodes the phase of matter. For superconductors, the non-Stückelberg sector vanishes, while for conductors, the leading-order terms are
\be
      I_{\rm oc}^{(\rm non)} = \int d^4 x \left(\sigma^{-1} \bm c_a \cdot (iT_0 \bm c_a - \p_t \bm c) - \xi \bm b\cdot\bm b_a \right). 
\ee
The resulting equations of motion for $I_{\rm oc}$ combine Maxwell’s laws with 
     $\bm E = \chi^{-1} \pp \rho + \mu\lam^2 \p_t \bm J - \sigma^{-1} \p_t\bm c ,$ and $ \bm B = - \xi \bm b -\mu\lam^2 \pp\times\bm J.$ 
At leading order in the derivative expansion, the second equation gives $\bm b=0$, while the first gives $\bm J = \sigma \bm E$. 

For vortex glasses, only $\bm b,\bm b_a$ are non-Stückelberg fields, yielding dynamics that enforce time-independent vortex-density solution $\bm b = \bm b_0(\bm x)$. Consequently
$\bm E = \mu \lam^2 \p_t\bm J,$ and $      \bm B - \lam^2 \pp^2 \bm B  = -\xi \bm b_0(\bm x).$

For chimeric conductors, only $\bm c,\bm c_a$ are non-Stückelberg fields, yielding action $I_{\rm cc} \equiv I_{\rm oc}|_{\xi =0}$, and equations of motion
\be
\bm E = \chi^{-1} \pp \rho + \mu\lam^2 \p_t \bm J - \sigma^{-1} \p_t\bm c ,\quad  \bm B = - \xi \bm b -\mu\lam^2 \pp\times\bm J.
\ee

{\it Chimeric insulator duality:}
The action for the chimeric insulator can be achieved by taking the $\sigma\to 0$ limit of $I_{\rm cc}$, which forces $\bm c_a=\p_t\bm c=0$. Plugging this back into the action, we find there is no dissipation so it can be expressed as an ordinary action 
     \be S_{\rm ci} = \int d^4 x\left( \ha \eps \bm E^2 - {1\ov 2\mu}\bm B^2  
     + \ha \mu\lam^2 (\pp\times \bm b)^2  + \pp\times\bm b\cdot \bm A \right).\ee

There is an equivalent description, which we can obtain via a duality transformation. First, promote $\pp\times\bm b \to \bm w$, where $\bm w$ is considered a fundamental field and introduce auxiliary scalar field $\psi$. Define the auxiliary action 
     \be S_{\rm aux} = \int d^4 x\left( \ha \eps \bm E^2 - {1\ov 2\mu}\bm B^2 + \ha \mu\lam^2 \bm w^2   + \bm w\cdot (\bm A - \pp\psi )  \right).\ee 
Notice that the equations of motion for $\psi$ yield $\pp\cdot\bm w = 0$. By the Helmholtz decomposition theorem, there exists vector $\bm b$ such that $\bm w=\pp\times\bm b$. Thus integrating out $\psi$ reproduces $S_{\rm ci}$, indicating these two actions are physically equivalent. To obtain the dual action, integrate out $\bm w$, yielding~\eqref{eq:chimins2}. 

{\it Global $U(1)$ order parameter:} 
The quadratic action for a superfluid is 
     $S_{\rm sf} = \int d^4 x ( \ha \chi \dot \th^2 - \nu (\pp \th)^2 ),$ 
where $\nu \to 1/\mu\lam^2$ in the superconductor action. Inspired by the continuum theory for the lattice model, the action for the chimeric superfluid is modified with auxiliary phase field~$\psi$: 
\be S_{\rm cs} = \int d^4 x \left( \ha \chi \dot \th^2 - \nu (\pp \th + \pp \psi)^2 \right).\ee  
The order parameter is 
$\vev{e^{i\th}} \propto \int d[\th\psi] e^{i S_{\rm cs}} e^{i\th}. 
$ 
Evaluating the path integral over $\psi$ first removes all gradient energy associated with $\th$. It immediately follows that $\vev{e^{i\th}} = 0$.

\end{document}